\documentclass[aps,prx,reprint,groupedaddress,doi=false,isbn=false,url=false,title=true,longbibliography,noeprint]{revtex4-2}
\usepackage{amsmath}
\usepackage{amssymb}
\usepackage{float}
\usepackage{graphicx}

\usepackage{xcolor}

\begin{document}

\title{Fast, hierarchical, and adaptive algorithm for Metropolis Monte Carlo simulations of long-range interacting systems}

\author{Fabio M\"uller}
\email{fabio.mueller@itp.uni-leipzig.de}
\author{Henrik Christiansen}
\email{henrik.christiansen@itp.uni-leipzig.de}
\author{Stefan Schnabel}
\email{stefan.schnabel@itp.uni-leipzig.de}
\author{Wolfhard Janke}
\email{wolfhard.janke@itp.uni-leipzig.de}
\affiliation{Institut f\"ur Theoretische Physik, Universit\"at Leipzig, IPF 231101, 04081 Leipzig, Germany}
\date{\today}

\begin{abstract}
  We present a fast, hierarchical, and adaptive algorithm for Metropolis Monte Carlo simulations of systems with long-range interactions that reproduces the dynamics of a standard implementation exactly, i.e., the generated configurations and consequently all measured observables are identical, allowing in particular for nonequilibrium studies.
  The method is demonstrated for the power-law interacting long-range Ising model with nonconserved order parameter and a Lennard-Jones system both in two dimensions.
  The measured runtimes support an average complexity $O(N\log N)$, where $N$ is the number of spins or particles.
  Importantly, prefactors of this scaling behavior are small, which in practice manifests in speedup factors larger than $10^4$.
  The method is general and will allow the treatment of large systems that were out of reach before, likely enabling a more detailed understanding of physical phenomena rooted in long-range interactions.
\end{abstract}

\renewcommand*{\eqref}[1]{Eq.~(\ref{#1})}

\maketitle
\section{Introduction}
\label{sec:intro}
The statistical physics of interacting $N$-body systems poses many important scientific problems that can be solved by analytic methods only approximately or in certain limits.
Therefore, they are nowadays often investigated by means of computer simulations which can be categorized into two main groups:
Molecular Dynamics (MD) simulations solve a system's equations of motion numerically and Monte Carlo (MC) simulations explore its phase space in a stochastic manner.
In both cases the interaction among the constituents of the system has to be taken into account.
For MD as forces and for MC as energy changes associated with random moves of the components.
With short-range interactions only a few other partners have to be considered while in the long-range case all the other constituents of the system are involved.
This severely limits the accessible system size $N$, since updating all constituents once naively requires $\sim N^2$ operations, usually labeled as complexity $O(N^2)$.
Since systems with long-range interactions are omnipresent in nature~\cite{eyink2006onsager,campa2009statistical,french2010long,levin2014nonequilibrium,campa2014physics,douglas2015quantum,neyenhuis2017observation,zhang2018long}, fast algorithms for their investigation are highly desirable. 
Consequently, there has been a lot of research proposing several methods addressing this inherent computational challenge.
\par
Two major classes of such methods are: \emph{i}) Methods based on splitting the evaluation of the potential into short- and long-range contributions, with one important example being the Ewald summation~\cite{perram1988algorithm} and \emph{ii}) hierarchical methods where groups of components are treated collectively such as the Barnes-Hut algorithm~\cite{barnes1986hierarchical}.
Algorithms from these two classes reduce the computational complexity to $O(N^{3/2})$ and $O(N \log N)$, respectively.
However, they have some disadvantages in certain situations (periodic vs. free boundaries, very large prefactors, control of systematic errors) and cannot all equally well be employed in MD and MC.
Only in MD, where all components progress synchronously, advanced algorithms based on fast multipole methods, particle mesh Ewald, and multigrid techniques which calculate all forces at the same time lead to even further reduced computational complexity.
Most of these studies focus on Coulomb interactions; for reviews see~\cite{sutmann2011fast,arnold2013comparison}.
In contrast, typically MC algorithms work asynchronously, i.e., they change only small parts of the system at a time.
There, these advanced methods cannot be used successfully, since calculating all interactions after each local update is wasteful even if done efficiently.
To achieve a similar improvement also for (asynchronous) Metropolis MC simulations of long-range systems, we here present a hierarchical adaptive algorithm that for the here considered systems reduces the computational complexity while maintaining a small prefactor without introducing any systematic errors.
The basic idea of our algorithm is the combination of the inverted Metropolis criterion with an adaptive tree-like spatial decomposition of the interaction energy.
Our algorithm reproduces exactly the same Markov chain as a traditional Metropolis implementation and can therefore be used as a one-to-one replacement, enabling in particular nonequilibrium studies as well.
\par
This is a major conceptual difference to MC methods that deviate from conventional Metropolis dynamics.
A prominent example are non-local cluster algorithms~\cite{luijten1995monte,fukui2009order,flores2017cluster} for the simulation of spin systems which can reduce complexity to $O(N)$, overcome critical slowing down and hence be more efficient for equilibrium studies close to criticality than any algorithm with local dynamics including the one presented here.
Furthermore, there is the rejection-free event-chain MC method for systems with continuous degrees of freedom~\cite{bernard2009event}.
It was first applied with great success to hard-sphere systems and has later been developed further to treat systems with general interactions~\cite{michel2014generalized,krauth2021event}.
It exploits that additive terms in the Hamiltonian transpose to factors in the Boltzmann weight and thus allow the application of a factorized Metropolis filter~\cite{hucht2009nonequilibrium,michel2014generalized}.
This idea has also been used in the recent development of the clock MC method~\cite{michel2019clock} which in contrast to event-chain MC is applicable to Ising systems as well and has a reported complexity of $O(N)$.
\par
In this study, we demonstrate the details of our algorithm by applying it to the nonconserved long-range Ising model (LRIM) with power-law decaying potential, where we focus on integrable interactions, and an off-lattice Lennard-Jones (LJ) system both in two dimensions.
In equilibrium, we find that the algorithm's performance is excellent and runtimes are in agreement with a complexity of $O(N \log N)$ for both systems.
For the LRIM which undergoes a second-order phase transition at the critical temperature $T_c$ we also consider two nonequilibrium processes: First quenches from a disordered configuration to a temperature $T<T_c$ into the ordered phase and secondly to $T_c$ itself.
The first case is referred to as phase-ordering kinetics where coarsening and aging phenomena~\cite{bray2002theory,full_book_puri,henkel2010non} occur.
In the latter setup critical aging~\cite{calabrese2005ageing,henkel2010non} can be investigated.
Such nonequilibrium processes are governed by local dynamics and consequently need to be modeled by local algorithms like the one presented here.
We will show that in all cases a significant speedup is achieved and will point out other systems where a similar performance can be expected.
\par
The rest of the paper is structured as follows:
In Section~\ref{sec:general_method} we first sketch the general procedure of our algorithm.
We then apply it to the LRIM in Section~\ref{sec:lrim} and the Lennard-Jones system in Section~\ref{sec:LJ}.
In Section~\ref{sec:applicability} we outline the applicability for a variety of other important (spin) models.
Finally, we conclude and give an outlook in Section~\ref{sec:conclusion}.

\section{Method}
\label{sec:general_method}
In this section we present a general formulation of the algorithm, with no reference to any system-specific properties.
\begin{figure*}
  \includegraphics[width=\textwidth]{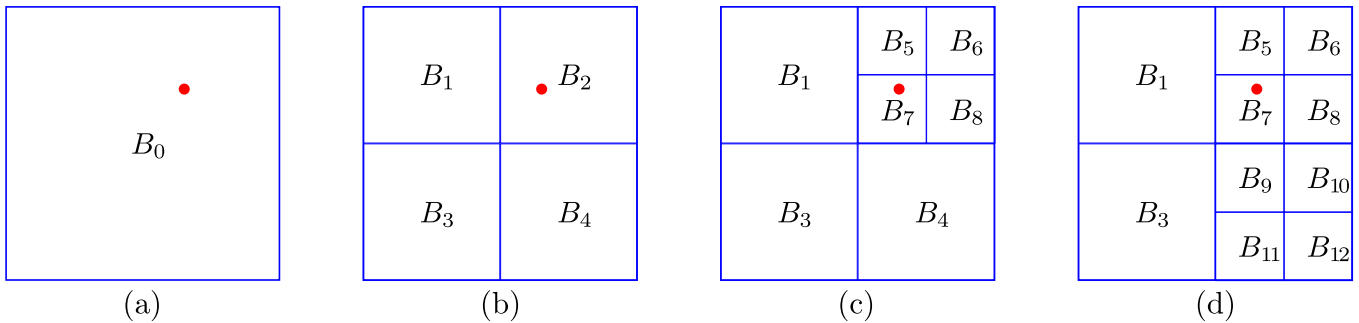}
  \caption{A visual sketch of an example progression for the decomposition of the interaction. The red dot marks the position of the component for which the update is proposed. The interacting boxes are enumerated in ascending order in which they are placed into the decomposition. From left to right always one box with high uncertainty is split into smaller boxes leading to more accurate bounds for $\Delta E$.}
  \label{fig:vis_sketch}
\end{figure*}
We consider a system of $N$ components  $\mathbf{q}= (q_1,\ldots,q_i,\ldots,q_N)$ where $q_i$ can for example stand for the spatial position $\mathbf{r}_i$, a binary Ising spin $s_i=\pm 1$, or other internal degrees of freedom (which may also be vector valued).
The components $q_i$ and $q_j$ interact via a symmetric pairwise potential $V_{i,j}=V_{j,i}$ such that the total energy reads
\begin{equation}
  \label{eq:total-energy2}
E = \frac12\sum_{i}\sum_{j \neq i} V_{i,j}.
\end{equation}
For a traditional Metropolis MC simulation with local dynamics at each step the update of a single randomly chosen component $q_i$ is proposed
\begin{equation}
  \label{eq:update-proposal}
  \begin{aligned}
  \mathbf{q}^\mathrm{old} &= (q_1,\ldots,q_i^\mathrm{old},\ldots,q_N) \rightarrow \\
  \mathbf{q}^\mathrm{new} &= (q_1,\ldots,q_i^\mathrm{new},\ldots,q_N)
  \end{aligned}.
\end{equation} 
The proposed update is accepted with the Metropolis probability,
\begin{equation}
  \label{eq:metropolis}
  P_\mathrm{acc}=\min \left( 1, e^{-\beta\Delta E} \right),
\end{equation}
where $\Delta E = E^\mathrm{new} - E^\mathrm{old}$ is the energy difference resulting from the update and $\beta$ is the inverse temperature.
This acceptance probability is then compared to a (pseudo-)random number $\rho\in[0,1)$.
If the calculated probability is larger than this random number the update is accepted and otherwise rejected.
As only one component $q_i$ of the system is updated we can write the change in energy of the whole system as,
\begin{equation}
  \label{eq:explicit_de}
  \Delta E =\sum_{j\neq i}  \left( V_{i,j}^\mathrm{new} -  V_{i,j}^\mathrm{old} \right). 
\end{equation}
Since in long-range interacting systems all constituents of the system interact with each other the calculation of $\Delta E$ requires the evaluation of $2(N-1)$ interactions.
A Metropolis MC sweep consisting of $N$ updates therefore has $O(N^2)$ complexity.
\par
The traditional way of performing a Metropolis simulation is to calculate $\exp(-\beta\Delta E)$ first and then comparing it to $\rho$.
That is an update is accepted only if 
\begin{equation}
  \label{eq:direct-metropolis}
  e^{-\beta\Delta E} \geq \rho
\end{equation}
is fulfilled.
Equivalently one may write
\begin{equation}
  \label{eq:inverted-metropolis-criterion}
  \Delta E \leq - \frac{\ln \rho}{\beta} \equiv \Delta E_{\mathrm{th}}
\end{equation}
and first draw the random number determining the threshold energy $\Delta E_{\mathrm{th}}$.
This shifts the decision about a proposed update to the problem of determining whether the actual energy difference $\Delta E$ involved in the update lies below or above $\Delta E_{\mathrm{th}}$.
In order to achieve that, $\Delta E$ does \emph{not} need to be known exactly.
Instead, it is enough to establish sufficiently narrow, rigorous bounds $\Delta E^{\min} \leq \Delta E \leq \Delta E^{\max}$.
The update is either accepted if the upper bound $\Delta E^{\mathrm{max}}$ is smaller than $\Delta E_{\mathrm{th}}$ or rejected if the lower bound $\Delta E^{\mathrm{min}}$ lies above $\Delta E_{\mathrm{th}}$.
Avoiding the direct calculation of $\Delta E$ can reduce the complexity and, thus, result in considerable speedups.
This procedure can easily be applied to other acceptance criteria such as, e.g., the Glauber acceptance rule, giving a different expression for $\Delta E_{\mathrm{th}}$.
\par
To construct the bounds $\Delta E^\mathrm{min/max}$ we perform a spatial decomposition of the simulation domain which is based on an extrinsic tree-like structure, in contrast to the intrinsic decomposition for self-avoiding walks~\cite{clisby2010accurate,clisby2010efficient} and polymers~\cite{schnabel2020accelerating}.
We note that any $d$-dimensional simulation box of linear size $L$ can be split into $2^d$ boxes of size $L/2$.
Of course, each of these boxes can again be split into $2^d$ boxes of size $L/4$ and so on.
This is repeated until each box contains no more than one constituent.
All theses boxes are thus automatically arranged hierarchically on a tree $\mathcal{T}$.
Inner nodes contain only the collective information needed for the estimation of the interaction, whereas within each leaf the single contained constituent is stored.
The construction of $\mathcal{T}$ has complexity $O(N)$ and rebuilding $\mathcal{T}$ completely at each update step would, therefore, be inefficient.
Instead we update $\mathcal{T}$ locally after an accepted update.
This requires $O(\log N)$ operations, since only the collective information of all the ancestor nodes of the leaf containing the updated component needs to be modified.
\par
This spatial decomposition of the simulation domain allows us to split the energy difference which follows from an update,
\begin{equation}
  \label{eq:grouped_de}
          \Delta E =\sum_{B \in \mathcal{D}} \Delta E_{B},
\end{equation}
where $\mathcal{D}$ is the set of currently selected, non-overlapping boxes (which may be of different size) covering the simulation space and $\Delta E_{B} = E^\mathrm{new}_{B}-E^\mathrm{old}_{B}$ is the exact change in energy contributed by the interaction with the constituents of box $B$.
Accordingly, if we can find strict lower and upper bounds $\Delta E^\mathrm{min/max}_{B}$ of $\Delta E_{B}$, we can establish bounds for the total energy change as well,
\begin{equation}
  \label{eq:general_total_bounds}
   \Delta E^\mathrm{min/max} =  \sum_{B \in \mathcal{D}}\Delta E^\mathrm{min/max}_{B}.
\end{equation}
General albeit not very tight bounds can be constructed by assuming that all constituents of a box are located at the points of minimal or maximal interaction, respectively.
\par
We aim at finding bounds $\Delta E^\mathrm{min/max}$ that are \emph{just} accurate enough to decide about the acceptance/rejection of a proposed update.
The general strategy is illustrated with an example progression of the decomposition $\mathcal{D}$ in Fig.~\ref{fig:vis_sketch}.
For easy visualization it is shown in two dimensions and for a non-moving component such as an Ising spin.
We start with the initial decomposition which is just the box containing the whole system $\mathcal{D} = \{B_0\}$, see panel (a) of the sketch in Fig.~\ref{fig:vis_sketch}, where the position of the component $q_i$ to be updated is marked by a red dot.
The bounds $\Delta E^\mathrm{min/max} = \Delta E_{B_0}^\mathrm{min/max}$ of this initial decomposition are in most cases not accurate enough to take the decision about the proposed update.
Thus we replace $B_0$ in $\mathcal{D}$ with the child boxes $B_{1},\ldots,B_{2d}$ of $B_0$ in $\mathcal{T}$, see Fig.~\ref{fig:vis_sketch}(b).
The bounds $\Delta E^\mathrm{min/max}$ are updated according to \eqref{eq:general_total_bounds}.
With each pair of bounds $\Delta E_{B}^\mathrm{min/max}$ there is an associated uncertainty $\Delta_{B}= \Delta E_{B}^\mathrm{max} - \Delta E_{B}^\mathrm{min}$ for each box in $\mathcal{D}$.
The sum of the uncertainties of the newly inserted child boxes is by construction always smaller or equal to the uncertainty of the removed parent box.
If the newly obtained bounds are not sufficiently narrow to reach a decision, further boxes are split until a decision can be made.
\par
In order to do so, a strategy is needed to decide which box to split next.
It is generally beneficial to split boxes which have a high uncertainty $\Delta_{B}$, since there the potential for improving the bounds is greatest.
A natural approach is, therefore, to always select the box with the greatest $\Delta_B$ for splitting.
Since searching an unordered set is computationally very expensive this would require to keep all elements of $\mathcal{D}$ strictly ordered with respect to their uncertainties and new boxes could only be added to $\mathcal{D}$ in logarithmic time $O(\log |\mathcal{D}| )$.
An overall faster way is to group boxes according to the integer part of their logarithmic uncertainties $\delta_B= \lfloor \log_2 \Delta_B \rfloor$ and always select some box from the non-empty set with the highest $\delta_B$.
The number of the necessary operations is now independent of the size of $\mathcal{D}$ and a significant computational overhead can be avoided this way.
\par
The process of sequential decomposition of boxes is sketched out in Fig.~\ref{fig:vis_sketch}.
The box $B_2$ in Fig.~\ref{fig:vis_sketch}(b) had a high uncertainty and was replaced by its four child boxes in $\mathcal{T}$, see Fig.~\ref{fig:vis_sketch}(c).
The next box to be split was $B_4$.
Such an adaptive spatial decomposition can be performed in most cases: on lattices, on graphs with different geometries and even in case of continuous spatial degrees of freedom although the nodes of $\mathcal{T}$ might not always correspond to simple square or cubic boxes.
% However, it is only useful for potentials which decay with distance.
% In fact, only when then the size of the boxes in the final decomposition grows with the distance from the updated component $q_i$, the majority of the interactions can be treated collectively and with this a significant speedup can be achieved.
% In the extreme case of mean field models such as the Sherrington-Kirkpatrick spin-glass it would be difficult (or even impossible) to come up with a set of efficient bounds which allow to treat multiple interactions at the same time.
% % Such a decomposition is only useful for potentials which decay with distance.
% % Only then the size of the boxes in the decomposition will grow with the distance from the updated component $q_i$, so that the majority of the interactions can be treated collectively.
\par
With this refinement protocol, we have effectively constructed a hierarchical, adaptive (spatial) decomposition of the interactions, which depends strongly on the current configuration, the energy difference due to the proposed update $\Delta E$, and the threshold energy $\Delta E_\mathrm{th}$.
After the decision has been made the next update starts again with $\mathcal{D} = \{B_0\}$, since if a different component is to be updated the final decomposition will likely be completely different.

\section{Long-Range Ising Model}
\label{sec:lrim}
\subsection{Model}
\label{sec:lrim-model}
As a concrete application we first consider the LRIM on a two-dimensional ($d=2$) $L\times L$ square lattice where every spin $s_i = \pm 1$ interacts with every other spin of the system via a decaying power-law potential.
This model has mostly been investigated for nonconserved order parameter, i.e., with varying magnetization $M=\sum s_i$, in numerous equilibrium studies~\cite{glumac1989finite,luijten1995monte,luijten1997,fukui2009order,horita2017upper,flores2017cluster} and has recently gained renewed interest in phase-ordering kinetics~\cite{christiansen2018,corberi2019one,corberi2019universality,christiansen2020aging,agrawal2020kinetics,christiansen2020zero}.
\par
The Hamiltonian of this system is given by
\begin{equation}
  \label{eq:hamiltonian}
 {\cal H}=-\frac12 \sum_{i}\sum_{j \neq i} J_{i,j}s_is_j,
\end{equation}
where for free boundary conditions the interaction couplings $J^\mathrm{FBC}_{i,j}$ decay with distance like
\begin{equation}
  \label{eq:interaction}
  J^\mathrm{FBC}_{i,j}= r(i,j)^{-(d+\sigma)}.
\end{equation}
Here, $r(i,j)=|\mathbf{r}(i,j)|$ is the Euclidean distance, $d$ is the spatial dimension and $\sigma>0$ is a tuneable parameter controlling the decay of the potential.
The self-interaction of the individual spins is set to zero, i.e., $J_{i,i}=0$.
To reduce finite-size effects we employ periodic boundary conditions, implemented via Ewald summation~\cite{ewald1921berechnung}, which takes all periodic images of the system into account.
For spins on fixed lattice positions the Ewald summation can be incorporated directly into the couplings~\cite{horita2017upper}:
\begin{equation}
  \label{eq:ewald_correction}
  J^\mathrm{PBC}_{i,j} = \sum^\infty_{\mu,\nu = -\infty} |\mathbf{r}(i,j)+\mu L \hat{e}_x+\nu L \hat{e}_y |^{-(d+\sigma)},
\end{equation}
which can be used in conjunction with the simple minimum image convention, avoiding a significant computational overhead.
\begin{figure}
  \centering
  \includegraphics[width=0.4\textwidth]{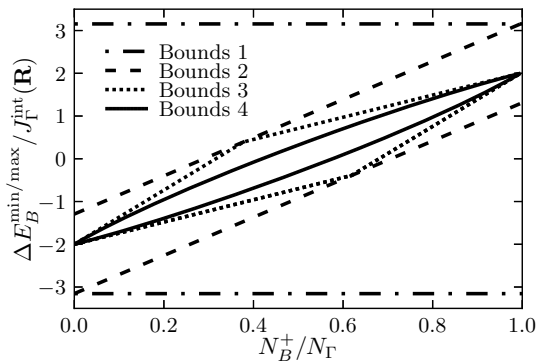}  
  \caption[Bounds for the interaction of a test spin with a box.]{Comparison of the different bounds $\Delta E_B^\mathrm{min/max}$ for the interaction of a test spin pointing in positive direction with a box $B$, normalized by the total interaction strength $J^\mathrm{int}_\Gamma$ defined in \eqref{eq:total_interaction_strength}.  A pair of bounds always consists of an upper and a lower bound which are kept in the same style.  $N_B^+$ is the number of spins pointing up and $N_\Gamma$ the total number of spins in the box. \textbf{Bounds 1} are given by $2N_\Gamma$ multiplied with the maximum interaction.  \textbf{Bounds 2} are refined by making use of the magnetization of the box. \textbf{Bounds 3} additionally employ the value of the interaction with the fully magnetized box $J_\Gamma^\mathrm{int}(\mathbf{R})$.  \textbf{Bounds 4} are the tightest bounds which can be obtained with the knowledge of the magnetization and the set of interactions involved in a spin-box interaction. For more information see text.}
  \label{fig:estimators}
\end{figure}
\par
For lattices with linear size $L=2^n$ with $n$ being a positive integer it is intuitive to decompose the lattice into smaller squares, which for the purpose of hierarchical access are arranged on a quad-tree.
Starting with the full lattice as the original box $B_0$ (cf. Fig.~\ref{fig:vis_sketch}) which encloses all the spins, we decompose the lattice into four boxes with half the linear size of the original box.
This is repeated until reaching the single spin level, where each box contains only a single spin.
We denote the level of decomposition as $\Gamma \in \{0,\ldots,n\}$ where $\Gamma=0$ corresponds to the full lattice and $\Gamma=n$ is the single spin level.
The linear size of a box is $L_{\Gamma}=2^{n-\Gamma}$ and the number of spins inside this box is $N_\Gamma=L^2_\Gamma=4^{n-\Gamma}$.

\subsection{Bounds $\Delta E_{B}^{\mathrm{min/max}}$ for the Spin-Box Interactions}
\label{sec:ising_estimators}
Proposing a flip of spin $s_i$, the contribution of the box $B$ to the energy difference can be written as
\begin{align}
  \label{eq:exact_interaction1}
  \Delta E_{B}
                &= 2 s^\mathrm{old}_i \sum_{j\in B} J_{i,j}  s_j,
\end{align}
where we identify the box $B$ with the set of the indices of the contained spins.
Here, we exploit that upon flipping $s_i$, $E_i^\mathrm{new}=-E_i^\mathrm{old}$.
For a lattice spin system with periodic boundary conditions the set of couplings which are involved in a spin-box interaction is always uniquely determined by the vector $\mathbf{R}$ from the spin to the center of the box and the size of the box $N_\Gamma$.
For the construction of the bounds $\Delta E_{B}^\mathrm{min/max}$, we need the minimal/maximal coupling in this set $J^{\mathrm{min/max}}_\mathrm{\Gamma}(\mathbf{R})$ (for monotonically decaying couplings they can be calculated in constant complexity $O(1)$ and can, therefore, be determined on the fly) and the integrated interaction 
\begin{equation}
  \label{eq:total_interaction_strength}
  J^\mathrm{int}_\Gamma(\mathbf{R}) = \sum_{j \in B} J_{i,j},
\end{equation}
which corresponds to the total interaction strength with a fully magnetized box $B$.
\par
For \textbf{Bounds 1} we do not discriminate between spins pointing up or down and only use the number of spins $N_\Gamma$ contained in the box.
It is clear that for each summand of~\eqref{eq:exact_interaction1} $-J_\Gamma^\mathrm{max}(\mathbf{R}) \leq J_{i,j} s^\mathrm{old}_is_j \leq J_\Gamma^\mathrm{max}(\mathbf{R})$ holds.
This allow us to formulate the following bounds for~\eqref{eq:exact_interaction1},
\begin{equation}
    \label{eq:boundaries_naive}
  \Delta E_{B}^\mathrm{max}=-\Delta E_{B}^\mathrm{min}=2N_{\Gamma}J^{\mathrm{max}}_\mathrm{\Gamma}(\mathbf{R}),
\end{equation}
which are plotted in Fig.~\ref{fig:estimators} as horizontal dashed-dotted lines as a function of $N_B^+/N_\Gamma$ for one example situation, where $N_B^+$ is the number of positive spins in the box which is related to the magnetization of the box $M_B= N_B^+-N_B^- = 2N_B^+ - N_\Gamma$.
The parameters for this example are linear lattice size $L=128$, decay exponent of the potential $\sigma=0.8$, distance $\mathbf{R}=(50.5,44.5)$ from the spin to the center of the box $B$, and box size $N_\Gamma=L^2_{\Gamma}=256$.
For other parameter values the curves in Fig.~\ref{fig:estimators} would look different, but the main features would remain unchanged.
Since these simple bounds do not make use of the box magnetization $M_B$ they are constant and much wider than the more refined bounds considered next. 
\par
In contrast, \textbf{Bounds 2} do depend on the magnetization $M_B$ in a box $B$.
To make use of $M_B$ we split the box $B$ into the sets of indices of spins pointing up $B^{+}$ or down $B^{-}$,
 \begin{equation}
  \label{eq:splitte_box}
   B =  B^{+} \cup B^{-},
 \end{equation}
 to rewrite \eqref{eq:exact_interaction1} as
\begin{equation}
  \label{eq:splitted_exact_interaction}
  \Delta E_{B}=2 s^\mathrm{old}_i\left(\sum_{j \in B^{+}}J_{i,j} - \sum_{j \in B^{-}}J_{i,j}\right).
\end{equation}
For each box of the spatial decomposition we keep track of the number of elements of $B^{\pm}$, i.e., the number of spins pointing up or down $N_B^\pm$. 
For the sums in \eqref{eq:splitted_exact_interaction} it follows that
\begin{equation}
  \label{eq:sum_bounds}
  N^\pm_{B}J^{\mathrm{min}}_{\Gamma}(\mathbf{R}) <\!\!\sum_{j \in B^\pm}\!\!J_{i,j} < N^\pm_{B}J^{\mathrm{max}}_{\Gamma}(\mathbf{R}).
\end{equation}
Assuming that $s^\mathrm{old}_i$ points into the positive direction, the rhs of~\eqref{eq:splitted_exact_interaction} becomes maximal if the first term is maximal and the second one is minimal and vice versa for its minimum.
This yields the following lower and upper bounds for the spin-box interaction:
\begin{align}
  \label{eq:boundaries_magnetization}
  \Delta E_{B}^\mathrm{min}&=2\left(N^+_{B}J^{\mathrm{min}}_{\Gamma}(\mathbf{R})-N^-_{B}J^{\mathrm{max}}_{\Gamma}(\mathbf{R})\right),\nonumber\\
  \Delta E_{B}^\mathrm{max}&=2\left(N^+_{B}J^{\mathrm{max}}_{\Gamma}(\mathbf{R})-N^-_{B}J^{\mathrm{min}}_{\Gamma}(\mathbf{R})\right).
\end{align}
If the test spin points in the negative direction, $N^+_{B}$ and $N^-_{B}$ have to be exchanged accordingly.
In this picture the upper bound $\Delta E_{B}^\mathrm{max}$ corresponds to the situation where all spins pointing in the same direction as the test spin are placed at the position of the strongest interaction and the spins pointing in the opposite direction at the spot of the weakest interaction.
For the lower bound $\Delta E_{B}^\mathrm{min}$ the positions are switched.
Looking at the dashed lines in Fig.~\ref{fig:estimators}, one clearly sees that these bounds indeed depend on the box magnetization $M_B$ and their uncertainty $\Delta_{B}$ is much smaller than for the rather loose \textbf{Bounds 1}.
\par
\begin{figure*}
  \newcommand{\mywidth}{0.27}
  \newcommand{\myspace}{0.1cm}
  \centering
  \includegraphics[width=\mywidth\textwidth]{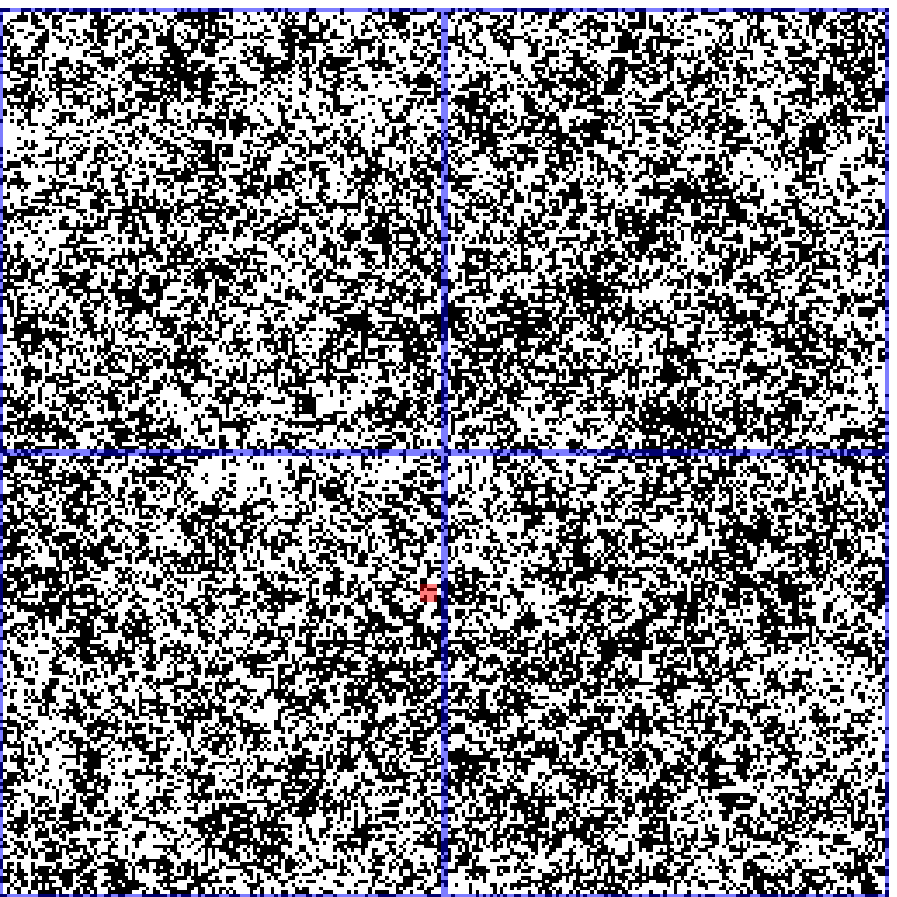} \hspace{\myspace} \includegraphics[width=\mywidth\textwidth]{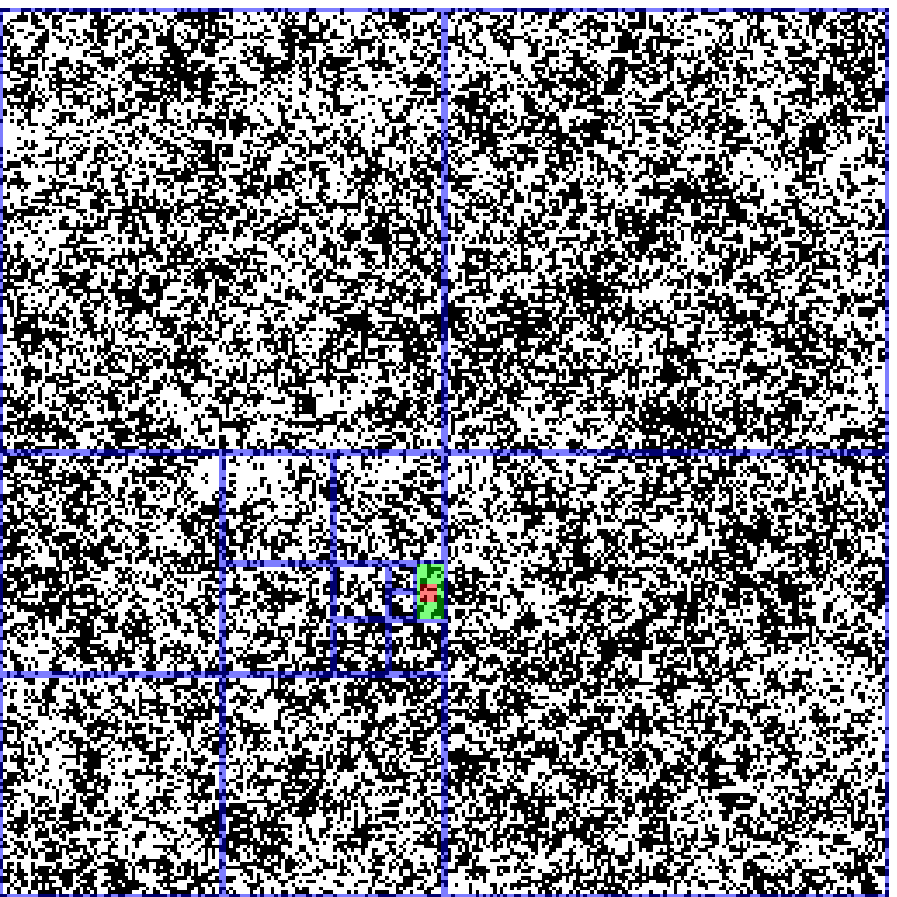} \hspace{\myspace} \includegraphics[width=\mywidth\textwidth]{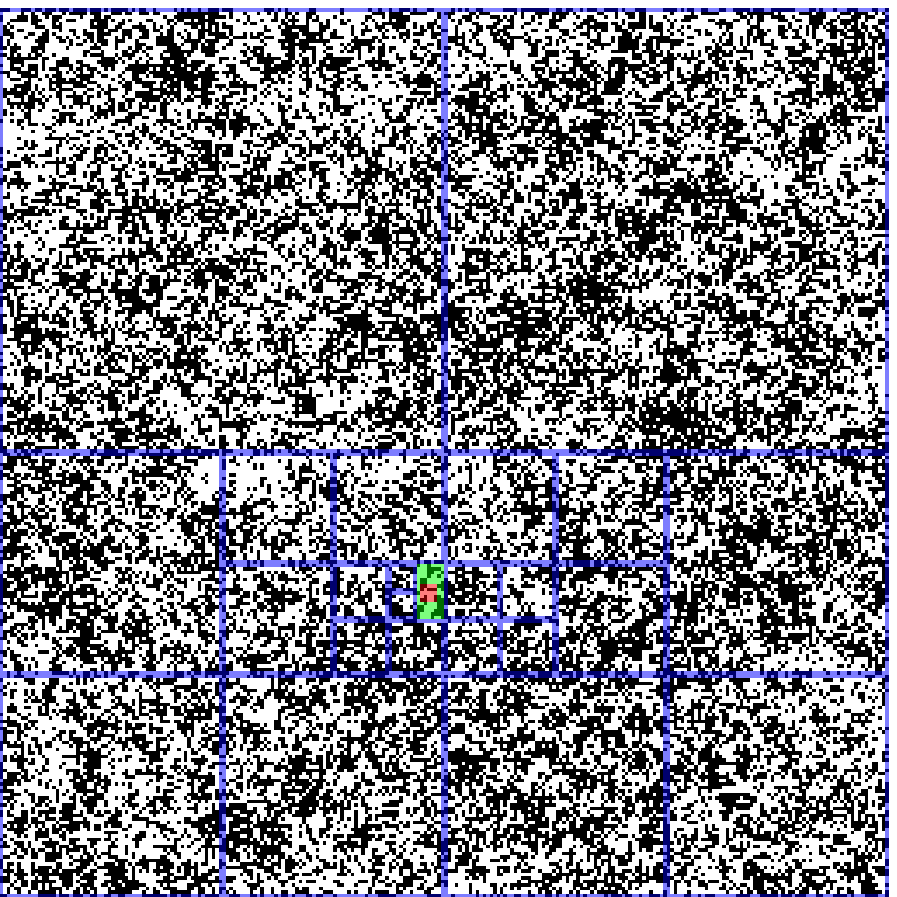} \\
  \vspace{0.2cm}
  \includegraphics[width=\mywidth\textwidth]{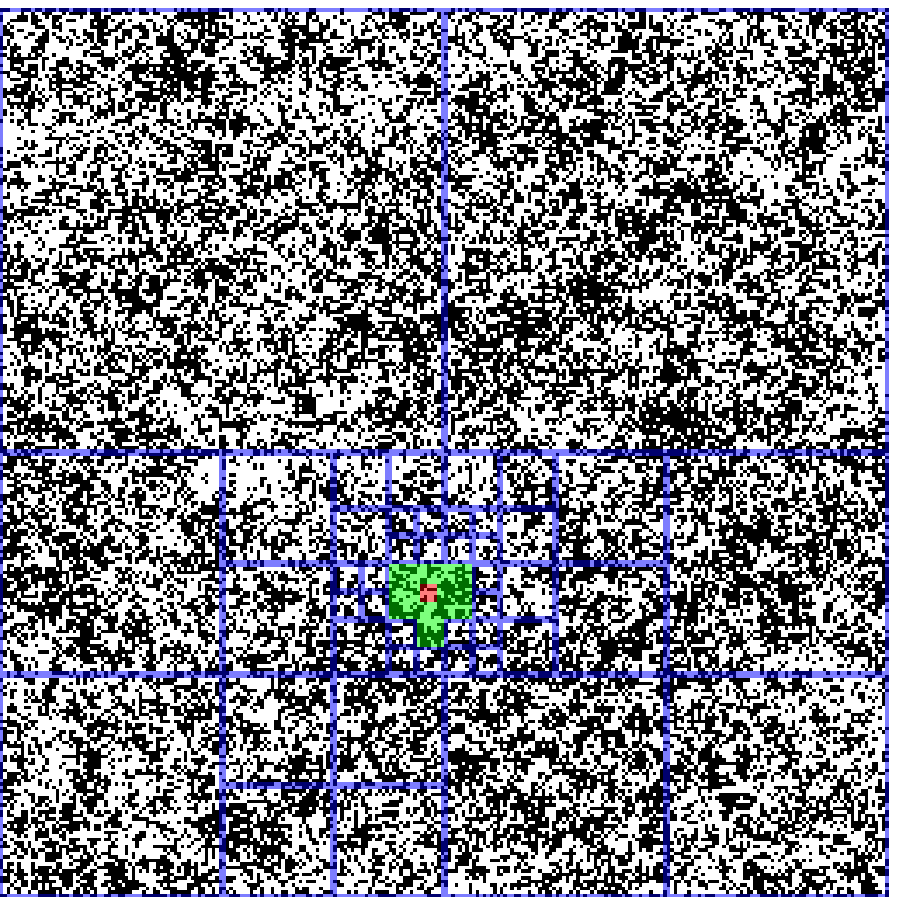} \hspace{\myspace} \includegraphics[width=\mywidth\textwidth]{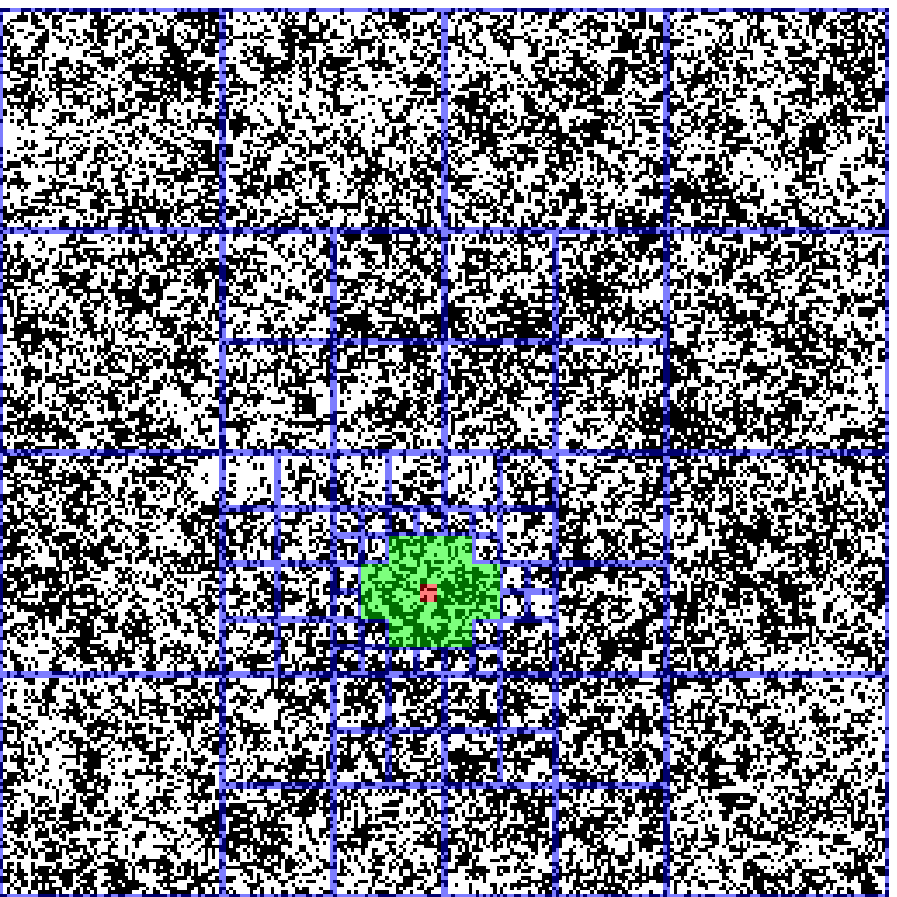} \hspace{\myspace} \includegraphics[width=\mywidth\textwidth]{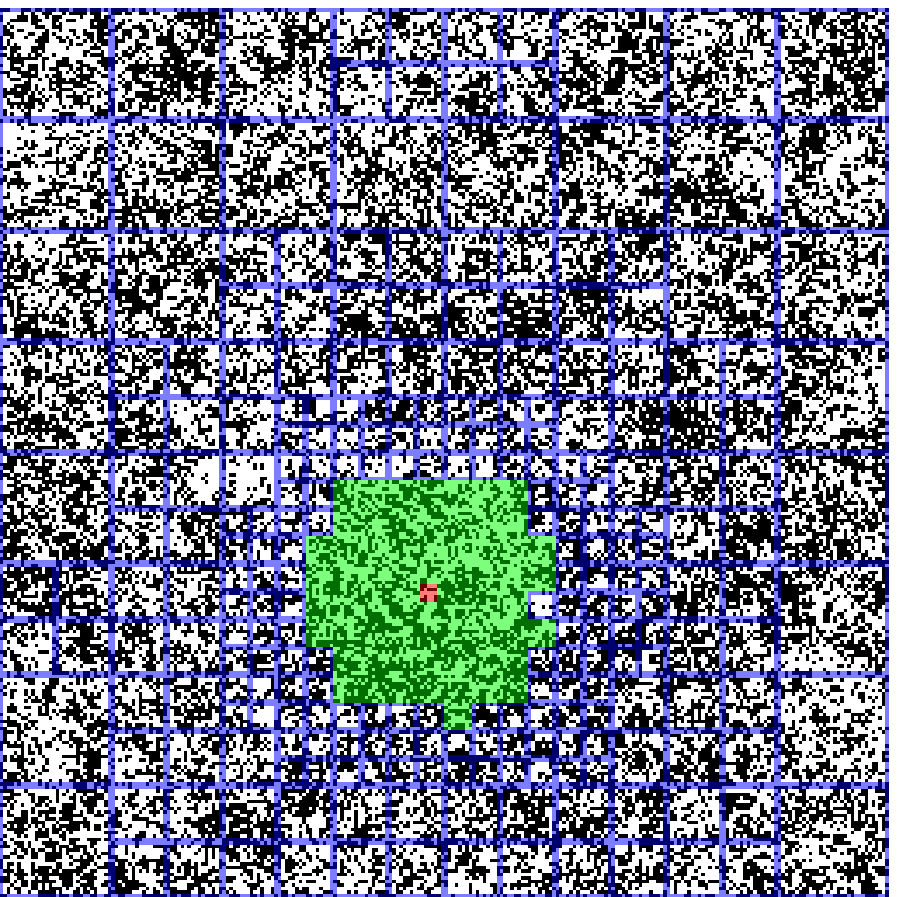}
  \caption{
    Example decomposition of the LRIM lattice with $\sigma=0.8$ and $L=256$ simulated in equilibrium at $T=T_c$.
    The spin which is to be updated is marked by the red dot close to the center of the box.
    The boundaries of the boxes are represented as blue lines.
    Going from top left to bottom right, the accuracy of the bounds of $\Delta E^{\mathrm{min}}$ and $\Delta E^{\mathrm{max}}$ increases with shrinking size of the boxes in the decomposition.
    Boxes of size $8\times8$ are broken up directly into $1 \times 1$ boxes  (green area); $2 \times 2$ and $4 \times 4$ boxes do therefore not occur.}
  \label{fig:decomposition}
\end{figure*}

For \textbf{Bounds 3} we use $J^{\mathrm{int}}_\mathrm{\Gamma}(\mathbf{R})$ from \eqref{eq:total_interaction_strength}.
Since the $J^{\mathrm{int}}_\mathrm{\Gamma}(\mathbf{R})$ do not depend on the spin configuration they can be precalculated.
The required computational effort is negligible compared to the typical simulation time, whereas the memory demands scale as $O (N \log N)$.
This can become challenging for systems that are significantly larger than the here considered system sizes.
Using $J^{\mathrm{int}}_\mathrm{\Gamma}(\mathbf{R})$, \eqref{eq:exact_interaction1} can be rewritten as
\begin{equation}
  \label{eq:exact_interaction_particle1}
   \Delta E_{B}=2 s^\mathrm{old}_i\left(J^\mathrm{int}_\Gamma(\mathbf{R}) - 2\sum_{j \in B^{-}}J_{i,j}\right).
 \end{equation} 
 This equation shows that the interaction of the test spin with a box with some arbitrary configuration can be seen as the sum of the interaction with the fully magnetized box where all spins point in the same direction as the test spin and twice the interaction of the spins inside the box which point in the opposite direction.
 Alternatively, the interaction can also be calculated using the interaction with the fully magnetized box with all spins pointing in the opposite direction of the test spin and adding twice the interaction with the spins parallel to the test spin,
 \begin{equation}
  \label{eq:exact_interaction_particle2}
   \Delta E_{B}=2 s_i^\mathrm{old}\left(2\sum_{j \in B^+}J_{i,j} - J^\mathrm{int}_\Gamma(\mathbf{R})\right).
 \end{equation}
 In order to derive two new pairs of bounds for the spin-box interaction we can again use \eqref{eq:sum_bounds}, inserting it into \eqref{eq:exact_interaction_particle1} and \eqref{eq:exact_interaction_particle2}.
 As both pairs of bounds are valid we can combine the two bounds for the minimum and the two bounds for the maximum by taking the tighter of the two and obtain (again for $s^\mathrm{old}_i = 1$),
 \begin{align}
  \label{eq:estimator_deviation}
  \Delta E_{B}^\mathrm{min}=2\max\big(&J^{\mathrm{int}}_\mathrm{\Gamma}(\mathbf{R})-2N^-_\mathrm{B}J^{\mathrm{max}}_\mathrm{\Gamma}(\mathbf{R}),\nonumber\\
                                      &2N^+_\mathrm{B}J^{\mathrm{min}}_\mathrm{\Gamma}(\mathbf{R})-J^{\mathrm{int}}_\mathrm{\Gamma}(\mathbf{R})\big),\nonumber\\
  \Delta E_{B}^\mathrm{max}=2\min\big(&J^{\mathrm{int}}_\mathrm{\Gamma}(\mathbf{R})-2N^-_\mathrm{B}J^{\mathrm{min}}_\mathrm{\Gamma}(\mathbf{R}),\nonumber\\
                                      &2N^+_\mathrm{B}J^{\mathrm{max}}_\mathrm{\Gamma}(\mathbf{R})-J^{\mathrm{int}}_\mathrm{\Gamma}(\mathbf{R})\big).
\end{align} 
These bounds are exact for fully magnetized boxes and still very accurate for almost fully magnetized ones (see the dotted lines in Fig.~\ref{fig:estimators}).
This greatly enhances the performance of the algorithm in presence of large magnetic domains, because the bounds of the boxes which fully lay inside a domain are very accurate and thus the decomposition can be coarser.
At very large distances $|\mathbf{R}| \gg L_\Gamma$ we find $J_\Gamma^{\mathrm{int}} \approx N_\Gamma (J_\Gamma^{\mathrm{min}} + J_\Gamma^{\mathrm{max}})/2$ so that the crossing of the pairs of bounds for the minimum or the maximum would occur at $N^{+}_{B} / N_\Gamma\approx 0.5$.
\par
Exploiting the fact that the set of $J_{i,j}$ can be sorted we can obtain even narrower \textbf{Bounds 4}.
For the second term in the brackets of \eqref{eq:exact_interaction_particle1} we replace the bounds from \eqref{eq:sum_bounds}, for which we assumed that all spins interact with $J^{\mathrm{min/max}}_\mathrm{\Gamma}(\mathbf{R})$, with the sum of the first $N^-_\mathrm{B}$ of the couplings sorted in ascending/descending order.
These are the tightest bounds which can be established using only the magnetization of the box and the set of couplings involved in the spin-box interaction.
These bounds are plotted in Fig.~\ref{fig:estimators} as solid lines and as the previous bounds can be calculated in $O(1)$ complexity if the above-mentioned sums are all computed and stored before the simulation.
The memory complexity of the algorithm using these bounds would scale as $O(N^2\log N)$, however, which limits the applicability for large system sizes.
Therefore, in the following we will employ \textbf{Bounds 3} which embody the best compromise between performance and memory requirements.
For the iterative refinement of the decomposition of the interaction we proceed as described in the general outline of the algorithm, with one modification.
It turned out to be beneficial to evaluate the interactions with small boxes via a direct summation of the spin-spin interactions since this is of comparable speed and has no uncertainty.

\subsection{An Example Decomposition}
\label{sec:example-decomposition}

In Fig.~\ref{fig:decomposition} we demonstrate the basic principle of our algorithm using \textbf{Bounds 3} by showing a single example snapshot and the corresponding spatial decomposition of the interaction for a simulation with $\sigma=0.8$ and $L=256$ at $T=T_c$ in equilibrium (the system is chosen to be relatively small, so that details can still be observed).
Here, the spin under consideration is positioned close to the center of the snapshot and is marked in red.
In the vicinity of the test spin -- the green-shaded region -- maximal resolution is reached, all the boxes are of size $1 \times 1$ and contain only a single spin each.
To ensure the possibility of arbitrarily precise estimation of $\Delta E$, interactions with these spins have to be considered exactly irrespective of which bounds are used, although we note that this is formally equivalent to the use of \textbf{Bounds 2-4}.
% The green shaded area indicates that interactions are resolved exactly, either because they are considered as spin-spin interactions or because the box has only spins pointing in one direction, for which the upper and lower bounds coincide.
% The interaction with the blue framed boxes enters as spin-box interactions.
\par
From top left to bottom right, the estimates of $\Delta E_{\mathrm{max}}$ and $\Delta E_{\mathrm{min}}$ are more refined and approach $\Delta E$.
As one can see, this is achieved by reducing the box sizes.
The decomposition adapts to the given configuration, i.e., regions which have a bigger influence on the decision are covered by smaller boxes.
The scenario in Fig.~\ref{fig:decomposition} required a rather fine-grained decomposition, but this is not representative and usually decisions can be made with a much coarser decomposition.
It is nonetheless a very illustrative example, since it nicely demonstrates the progression of the algorithm.

\subsection{Analysis of the Runtimes}
\label{sec:runtime_analysis}
The speed of our algorithm strongly depends on the state of the simulation, i.e., the current configuration, the choice of the spin to be updated, the temperature, etc.
The decision becomes harder the closer the involved change in energy $\Delta E$ is to the threshold energy $\Delta E_\mathrm{th}$.
For the extreme case of $\Delta E =\Delta E_\mathrm{th}$ the interaction of the updated spin with all other individual spins has to be evaluated exactly, implying a \emph{worst case} complexity of the algorithm of $O(N^2)$ per sweep.
In the case of high temperatures, one has $\Delta E_\mathrm{th} \rightarrow \infty$, which means that the actual change in energy of the proposed update becomes irrelevant, so that the initial bounds are sufficiently narrow to accept it.
Only the updating of the tree $\mathcal{T}$ has to be performed, which yields the \emph{best case} complexity $O(N \log N)$.
\par
It is not straightforward to predict the \emph{average} complexity from the algorithm's design alone, so that, in the following, we will record the resulting runtimes $\tau$ per sweep for different physical settings.
For all measurements we take care of minimizing possible hardware-related influences.
To be on the safe side and to not draw any wrong conclusions in our analysis, we have nonetheless assumed an error of $10\%$ on our estimated runtimes to account for remaining errors.
\subsubsection{Equilibrium}
\begin{figure}
  \centering
  \includegraphics[width=0.4\textwidth]{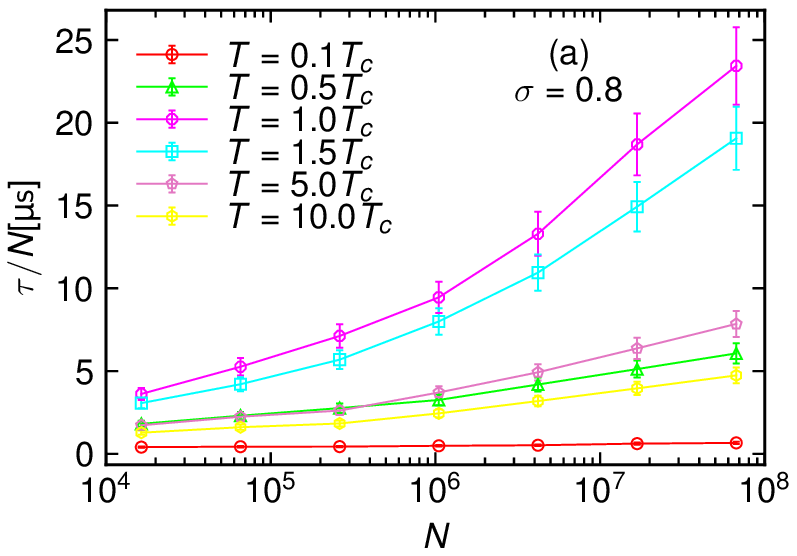}\\
  \includegraphics[width=0.4\textwidth]{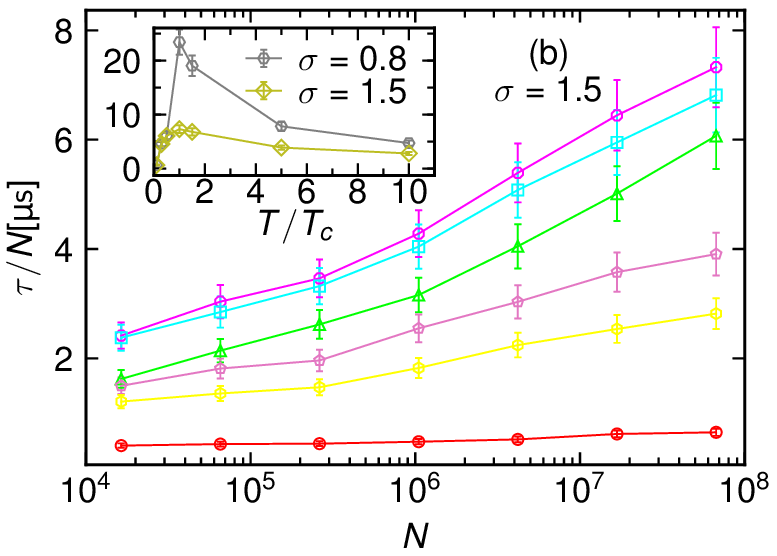}
  \caption{Runtime per spin update $\tau/N$ versus system size $N$ on a semi-log scale for (a) $\sigma=0.8$ and (b) $\sigma=1.5$ and several (equilibrium) temperatures $T$.
  In the inset of (b) we show the combined data for both values of $\sigma$ and the biggest system size $L=8192$ as a function of $T$, demonstrating the maximum around $T_c$.}  
  \label{fig:runtime_equibrium_nonconserved}
\end{figure}
For equilibrium simulations, the computational cost can relatively straightforwardly be extracted from rather short runs, which allows us to consider a broad temperature range.
A further advantage is that the runtimes only weakly depend on the initial conditions and can be averaged over the full run after the equilibration.
The benchmarking was performed on a single hardware configuration: Single socket motherboard equipped with an Intel Core i5-8500T CPU and 16GB DDR4-2667 dual-channel RAM.
The algorithm was implemented in C++17 and compiled using GCC 8.3.
Since modern processors do not run using a constant frequency and use speculative execution, results of benchmarks can fluctuate.
This is especially a problem, if the compute nodes are occupied by other tasks.
Therefore, we have taken care to exclusively run a single simulation at a time per compute node in order to minimize possible fluctuations.
\par
\begin{figure}
  \centering
  \includegraphics[width=0.4\textwidth]{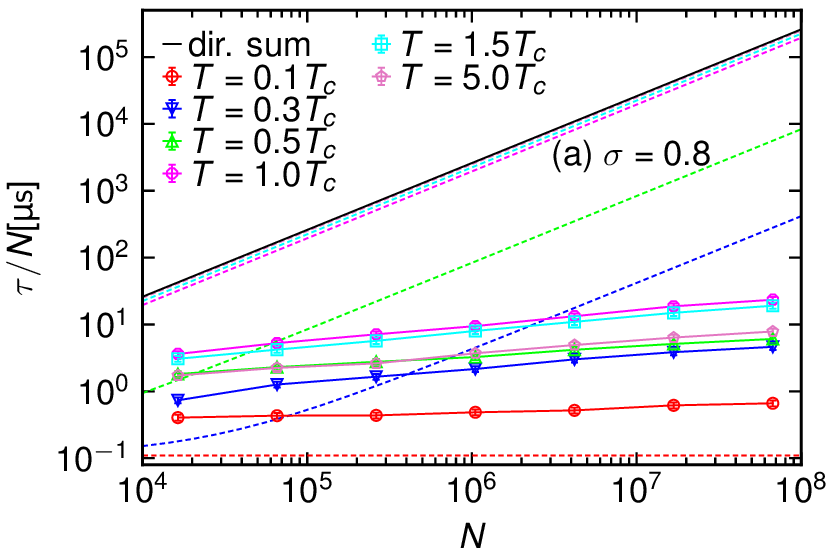}\\
  \includegraphics[width=0.4\textwidth]{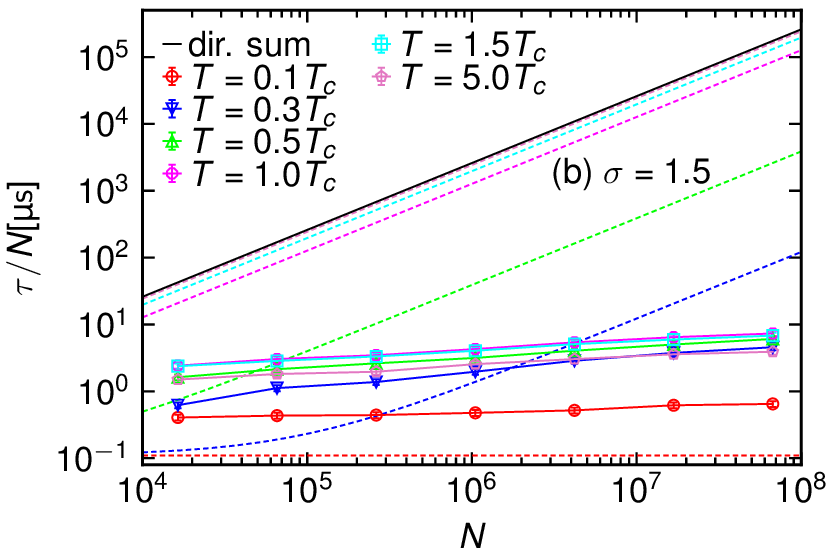}
  \caption{Runtimes per spin update $\tau/N$ versus system size $N$ plotted on a log-log scale for (a) $\sigma=0.8$ and (b) $\sigma=1.5$ and several (equilibrium) temperatures $T$.
    As dashed lines in the same color as the original data points, we have included the calculated runtimes for an effective field simulation.
  Also included is the runtime obtained from a naive Metropolis MC simulation using direct summation.}  
  \label{fig:runtime_equibrium_nonconserved_comparison_eff_field}
\end{figure}

In Figs.~\ref{fig:runtime_equibrium_nonconserved}(a) and (b) the runtimes per spin update $\tau/N$ (in units of $\mu s$) for different fractions of $T_c$ in dependence of the system size are presented in a semi-log plot for (a) $\sigma=0.8$ and (b) $\sigma=1.5$.
In both cases the growth of the runtimes crosses over to linear behavior on the semi-log scale irrespective of the temperature $T$, in a manner compatible with $O(N \log N)$ complexity.
This we deem very plausible considering the hierarchical progression of the algorithm through the use of a tree.
Based on this data (and Fig.~\ref{fig:runtime_equibrium_nonconserved_comparison_eff_field} where we plot the runtimes on a log-log scale), a power-law complexity $O(N^{1+\alpha})$ with a small exponent $\alpha$ cannot, however, be completely ruled out.
In order to corroborate either of the two hypotheses, significantly larger systems need to be considered.
While a dedicated investigation of moderately larger system sizes could still be feasible in principle, significantly larger sizes are out of reach with the hardware used in this study, due to the large memory requirements, and a final assessment of the asymptotic scaling has to be left for future studies.
\par
We find that there is a clear dependence of the runtimes on the simulation temperature, which is also visualized in the inset of Fig.~\ref{fig:runtime_equibrium_nonconserved}(b).
This can be understood qualitatively from the considerations made before:
The maximum close to the critical temperature stems from the fact that the average threshold energy $E_\mathrm{th}$ is on average close to the actual energy difference $\Delta E$ involved in the proposed spin flips, which makes a fine decomposition of the interaction necessary.
In the case of very high temperature one has $\Delta E_\mathrm{th} \gg \Delta E$ and the actual change in energy of the proposed spin flip becomes irrelevant.
At low temperatures an opposite mechanism is at work.
Since $\Delta E_\mathrm{th} \rightarrow 0$ almost only spin flips are accepted that do not increase the energy, but a typical configuration at these temperatures is (nearly) ordered.
Thus, a single spin flip on average has $\Delta E \gg \Delta E_\mathrm{th}$ so that also this decision can be made with loose bounds.
\par
The comparison most relevant is likely the achieved speedup factor, which can be visually appreciated from Figs.~\ref{fig:runtime_equibrium_nonconserved_comparison_eff_field}(a) and (b), where we show the runtimes on a log-log scale in comparison to an effective field simulation~\cite{christiansen2018} (dashed colored lines in the same color as the original data points) and a direct summation (solid black line) of all interactions (for the sake of better comparability, here we have replaced $T=10 T_c$ by $T=0.3 T_c$).
The effective field approach uses a relatively simple storage trick to save many calculations whenever an update is \emph{not} accepted.
This yields a massive speedup whenever low acceptance rates are encountered, but requires the same number $O(N)$ of operations for an accepted update as in the direct summation, resulting in the same computational complexity $O(N^2)$~\footnote{For the effective field simulation times, we have performed simulations having different acceptance rates $\alpha$ (corresponding to the different temperatures) for a range of smaller system sizes and phenomenologically fitted the resulting runtimes with the function $R(\alpha)=AN\alpha+B$ giving the runtime per update, where $N$ is the number of spins, and $A$ and $B$ are free fitting parameters. It is thus assumed explicitly that the complexity is $O(N^2)$. Further, it is recognized that there is a constant, system size independent term $B$ involved. All the runtimes for the effective field are not measured directly but calculated from the system size and the acceptance rates measured during the simulation with the new algorithm.}.
Thus, the resulting runtimes are strongly dependent on the acceptance rate of the simulation, which means that this approach is especially fast at low temperatures where the acceptance rates are low~\cite{burovski2019acceptance}.
At high temperatures, many more proposed updates are accepted, which gives the effective field approach only a minor advantage over a direct summation.
At the critical temperature (for which our algorithm has the highest runtime), we can report a speedup factor of $\approx 5500$ for $\sigma=0.8$ and $\approx 12300$ for $\sigma=1.5$ compared to the effective field approach ($\approx 11000$ resp. $\approx 35000$ compared to direct summation).
For a temperature below $T_c$, e.g., for $T=0.5T_c$ we observe a speedup of $\approx 1700$ for $\sigma=0.8$ and $\approx 500$ for $\sigma=1.5$ ($\approx 40000$ resp. $\approx 43000$ compared to direct summation).
For increasing system size all these factors grow steadily.
For each fixed temperature there will be a crossover size above which our new algorithm is faster than the effective field method.
From Fig.~\ref{fig:runtime_equibrium_nonconserved_comparison_eff_field} one reads off that the crossover for $T = 0.3T_c$ occurs at $N \approx 10^6$  for the considered values of $\sigma$ and respectively $N \approx 10^5$ for $T=0.5T_c$.
For larger values of $T$ the crossover happens already for much smaller systems while for the very low temperature $T=0.1T_c$ the crossover cannot yet be observed for the considered system sizes of up to $N=10^8$ due to the extremely low acceptance rates ($\approx 10^{-6}$) and hence a very small prefactor of the runtimes of the effective field approach.
\par
For investigations of the phase transition, i.e., for equilibrium simulations in the proximity of $T_c$, our method, like any other local algorithm, is not a serious contender, at least when non-local cluster algorithms are available as for the Ising model.
In such cases, its main field of application are nonequilibrium scenarios.

\subsubsection{Nonequilibrium}
We focus on two cases: Quenches from a disordered start configuration \emph{i}) to a temperature substantially below the critical temperature or \emph{ii}) to the critical temperature.
To mimic a physical evolution, in these cases only local dynamics that preserve the dynamical properties of the system may be used.
Non-local update schemes, including cluster algorithms, are not allowed, which makes these scenarios the prime field of application for our algorithm.
In the first case, the system undergoes an ordering process and consequently the dynamics of growth of ordered structures in the system is of interest, both from coarsening and aging perspective involving single- and two-time quantities, respectively~\cite{bray2002theory,henkel2010non,full_book_puri}.
The physical properties of this model during this process have recently been investigated in Refs.~\cite{christiansen2018,christiansen2020aging,agrawal2020kinetics,christiansen2020zero} and are not part of the discussion here.
\begin{figure*}
  \centering
  \includegraphics[width=0.4\textwidth]{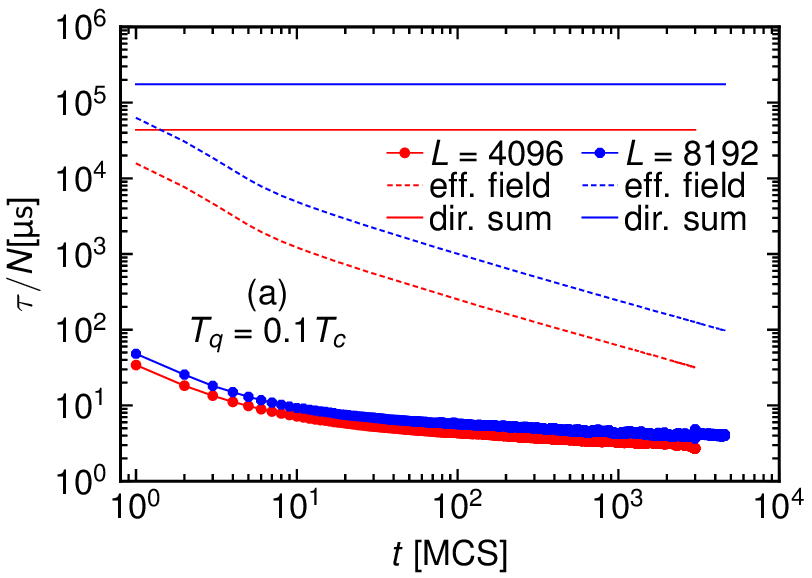}
  \hspace{0.066\textwidth}
  \includegraphics[width=0.4\textwidth]{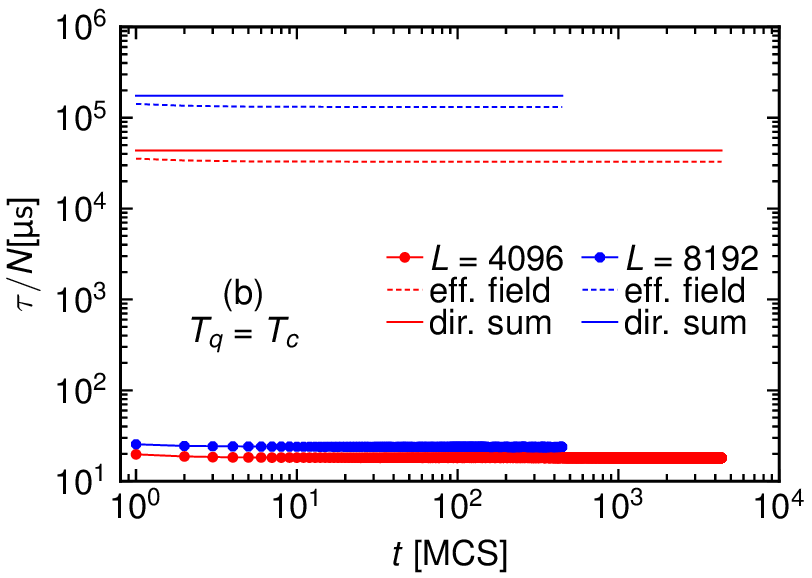}
  \caption{Runtime per spin update $\tau/N$ versus simulation time $t$ for $\sigma=0.8$ in quenches from a disordered start to (a) $T_q=0.1T_c$ and (b) $T_q=T_c$ on large lattices with $L=4096$ and $8192$.
    In both plots, we present as solid lines with markers the runtimes obtained by our algorithm.
    The dashed lines in the same color show the runtimes calculated for the effective field (eff. field) algorithm and the solid lines are the runtimes of the (naive) direct summation (dir. sum).
    }
  \label{fig:runtime_evolution}
\end{figure*}
In Fig.~\ref{fig:runtime_evolution}(a) we present the time dependence of the runtime of our algorithm per spin update $\tau/N$ (in units of $\mu s$) as a function of simulation time for a phase-ordering quench with $\sigma=0.8$ to $T_q=0.1T_c$ for large systems of linear sizes $L=4096$ and $8192$.
Also shown, for sake of comparison, are the runtimes for the effective field method and a direct summation of the interactions.
We could quench to arbitrary temperatures with our new algorithm, but choose $T_q=0.1T_c$ to have the same setting as in Ref.~\cite{christiansen2018}.
\par
We observe that the time needed per update $\tau/N$ is strongly dependent on how far the system has proceeded in its ordering process for our algorithm and, even more, for the effective field approach.
Since the temperature in our algorithm is set to a low temperature $T_q=0.1T_c$, we typically draw threshold energies $E_{\mathrm{th}}$ comparatively close to zero, i.e., spin flips which significantly increase the energy are usually not accepted.
At the start of our simulation the configuration is completely disordered, and for many proposed spin flips $\Delta E \approx 0 \approx \Delta E_\mathrm{th}$, so that $\Delta E$ has to be known rather accurately.
In the course of the simulation, when the configurations are already partly ordered, proposed spin flips in domains have typically a large $\Delta E$ and can mostly be rejected with very loose bounds.
The decision is more involved for spins positioned at domain boundaries, where $\Delta E$ is typically much closer to zero, the interactions often have to be resolved in more detail, and the probability of acceptance is higher.
With growing domains fewer spins are situated at the domain boundaries so that the average acceptance rate decreases, and the effective field simulations become faster.
Ultimately, the runtimes of both methods will reach their respective equilibrium runtimes at $T_q$.
\par
Over the full process, our new approach is significantly faster than the already very fast effective field method, resulting in a $\approx 100$ times faster total runtime until finite-size effects are reached.
A direct summation becomes in these cases prohibitively expensive (a factor of $\approx 40000$ slower than our new algorithm), and cannot be used to simulate systems of this size.
This factor will grow significantly for increasing system size since the runtimes of the two algorithms scale differently.
\par
The second case of interest is critical aging, i.e., the behavior of two-time quantities during quenches from a disordered starting configuration to the critical temperature~\cite{calabrese2005ageing,henkel2010non}.
With some modifications, such as a small initial magnetization, it is also possible to investigate short-time dynamics during such processes~\cite{janssen1989new}, but we here focus on the completely disordered start.
We present in Fig.~\ref{fig:runtime_evolution}(b) the obtained runtimes per spin update $\tau/N$ (again in units of $\mu s$) in our simulation with $\sigma=0.8$, where we use the same notation as in (a) for the different methods.
Here, the runtimes both for the new algorithm and the effective field approach remain more or less constant throughout the whole simulation, although the system has not yet reached equilibrium.
In both cases only a small initial decay of the runtimes is visible.
Here, the advantage of the effective field simulation over a direct summation is relatively small, since the acceptance rates are of the order of 1.
Albeit the equilibrium simulation close to the critical temperature is also one of the most difficult situations for our algorithm, it nonetheless produces much smaller runtimes than the effective field approach and the direct summation.
In this scenario the runtimes are close to those found in the equilibrium simulations.
We find a speedup of $\approx 6000$ compared to the effective field approach and $\approx 8000$ to the direct summation, allowing for the investigation of this process for the presented system sizes, which was entirely out of reach before.
A similar acceleration is also expected for other nonequilibrium simulations with comparably high acceptance rates.
\par
Another local algorithm for long-range interacting spin systems is the clock MC method~\cite{michel2019clock} based on the factorized Metropolis filter~\cite{hucht2009nonequilibrium,michel2014generalized}.
It, too, is potentially applicable in the scenarios discussed above.
Yet, so far it has only been applied to \emph{disordered} long-range interacting spin systems in \emph{equilibrium}. 
We have implemented this method for the ferromagnetic LRIM and tested it for the above nonequilibrium setup.
In its basic form where spin-spin interactions are considered individually, we find that the times of crossover to the asymptotic scaling behavior become prohibitively large for quenches to low temperatures.
This is due to drastically reduced acceptance rates of the factorized Metropolis filter, when compared to conventional Metropolis dynamics.
In the framework of this method there is, however, the possibility to treat multiple factors collectively.
This shifts the dynamics towards traditional Metropolis, increasing the acceptance rate, but also the computational effort.
For each physical setting (i.e., combination of $T$, $L$, $\sigma$, \ldots) a different grouping of spins may yield the best performance.
A detailed analysis and comparison is beyond the scope of this study and will be presented elsewhere~\cite{InPrep}.

\section{Lennard-Jones System with Full Interaction Range}
\label{sec:LJ}
To highlight the power of our algorithm also for continuous degrees of freedom, we finally demonstrate the applicability of its general concept to a LJ system~\cite{frenkel2001understanding} with potential
\begin{equation}
  V_{LJ}=4\epsilon \left[ \left( \frac{\sigma}{r_{ij}}\right)^{12} - \left(\frac{\sigma}{r_{ij}}\right)^6\right].
\end{equation}
Here, we keep the full interaction range, i.e., do not truncate (and shift) the potential at the often employed cut-off $r_c=2.5\sigma$.
It is well known that, e.g., the critical temperature and critical density of a LJ system do depend on $r_c$~\cite{smit1991vapor,smit1992phase,trokhymchuk1999computer}.
\par
We consider $N$ interacting particles in a volume $L^d$ whose linear extent $L$ can be adjusted to yield the desired density $\rho=N/L^d$.
Periodic boundary conditions are applied which, due the fast decay of $V_{LJ} \propto - r_{ij}^{-6}$ can easily be realized by the minimum image convention, i.e., in this application Ewald summation is not necessary.
\par
For the here presented simulations, we used the most general bound introduced in Sec.~\ref{sec:general_method}, i.e., we virtually collect all particles of a box at the points of minimal or maximal interaction, respectively.
This is analogous to \textbf{Bounds 2} introduced in Sec.~\ref{sec:ising_estimators} for the LRIM.
In equilibrium simulations of particles the proposed MC moves can be freely chosen.
Here we perform 90\% local displacements within radius $r=\sigma$ and 10\% nonlocal moves where the particle's potential new position is chosen randomly in the whole simulation box.
\begin{figure}
  \centering
  \includegraphics{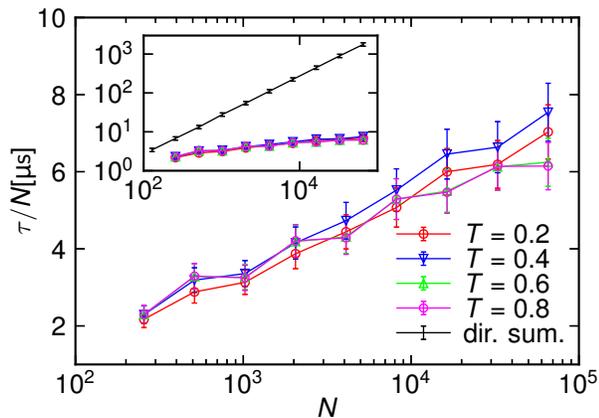}\\
  \caption{Runtime per update $\tau/N$ versus system size $N$ plotted on a log-normal scale for a Lennard-Jones system with untruncated interactions at constant density $\rho=0.35$.
    The inset shows the same data on a log-log scale together with the runtimes obtained from a standard Metropolis simulation using a direct summation of all interactions.
  }  
  \label{fig:LJ}
\end{figure}
In Fig.~\ref{fig:LJ} we show for $d=2$ the runtimes per update $\tau/N$ for different $N$ but fixed density $\rho=0.35$ and varying temperature $T$, covering both the (oversaturated) vapor and vapor-liquid phase.
We find clear evidence for $O(N \log N)$ complexity.
The runtimes appear largely independent of the temperature, which is in contrast to the results for the LRIM in equilibrium.
For smaller densities, we generally find faster runtimes for the same number of particles.
An implementation for $d=3$ is straightforward, too, and will be presented elsewhere~\cite{ToBePublished2}.

\section{Applicability \& Limitations}
\label{sec:applicability}
Although in this paper we only presented in detail the application of the algorithm to two different models, it can be used for other lattice spin and off-lattice particle systems as well.
Of course, we are neither restricted to two spatial dimensions nor to hypercubic lattices.
The method is versatile; besides the classical Ising model general $O(n)$ spin models can be treated efficiently in a very similar manner.
In particular, systems with quenched disorder~\cite{Young1997} are not excluded:
Random field models~\cite{Bray1985} are trivially accommodated within the framework introduced above since the extra field simply enters as an offset to $\Delta E$.
Another class of models are site-diluted spin systems which model crystal defects through unoccupied lattice sites~\cite{Harris1974,[][{, and references therein.}]Kazmin2020}.
To treat for instance a site-diluted Ising system with our algorithm, modified \textbf{Bounds 3} can be used,
\begin{align}
  \label{eq:estimator_deviation_diluted}
  \Delta E_{B}^\mathrm{min}=2\max\big(&J^{\mathrm{int}}_\mathrm{\Gamma}(\mathbf{R})-(2N^-_\mathrm{B}+N_B^0)J^{\mathrm{max}}_\mathrm{\Gamma}(\mathbf{R}),\nonumber\\
                                      &(2N^+_\mathrm{B}+N_B^0)J^{\mathrm{min}}_\mathrm{\Gamma}(\mathbf{R})-J^{\mathrm{int}}_\mathrm{\Gamma}(\mathbf{R})\big),\nonumber\\
  \Delta E_{B}^\mathrm{max}=2\min\big(&J^{\mathrm{int}}_\mathrm{\Gamma}(\mathbf{R})-(2N^-_\mathrm{B}+N_B^0)J^{\mathrm{min}}_\mathrm{\Gamma}(\mathbf{R}),\nonumber\\
                                      &(2N^+_\mathrm{B}+N_B^0)J^{\mathrm{max}}_\mathrm{\Gamma}(\mathbf{R})-J^{\mathrm{int}}_\mathrm{\Gamma}(\mathbf{R})\big),
\end{align}
where $N_B^0$ is the number of vacancies in box $B$.
This opens a way to treat $q$ state Potts models as well~\cite{[][{, and references therein.}]Majumder2018}, where the components which are inert to both the old and the newly proposed state would be treated as vacancies.
Now, we need to store the population of each of the $q$ spin states for all boxes.
\par
Random field and site dilution are forms of disorder that can easily be managed since they affect individual spins and their interaction with the environment as a whole.
More challenging are models where disorder manifests as variation of the interaction of pairs of spins such as systems with bond dilution~\cite{Kole2022} or the Edwards-Anderson spin-glass model~\cite{Edwards1975,Katzgraber2003}.
Here, the order parameter is not as closely related to the spin-box interaction energy as the magnetization in the case of the pure Ising model.
This implies that it is difficult to formulate an estimator similar to \textbf{Bounds 3} or \textbf{Bounds 2}.
Also in this more difficult case we have checked that a reduction in complexity is achieved in simulations using the basic \textbf{Bounds 1}, although the speedup is less pronounced as compared to the pure Ising case.
Another large class of problems are spin systems with competing interactions, e.g., antiferromagnetic short-range and ferromagnetic long-range interactions or vice versa~\cite{Giuliani2006,horowitz2015phase}.
Here, one could for example evaluate the short-range interactions directly and treat the long-range interactions using our algorithm.
\par
The LJ system considered above can easily be generalized to two-body potentials of other functional forms and extended to multi-species systems.
Depending on the specific details of the different interaction between the particles many scenarios can be imagined.
A general approach would be to use a separate decomposition of the system for each particle type.
\par
Yet, the question about possible limitations of the algorithm arises.
Cases where it might not be successful are systems with interactions that do not decay sufficiently fast (and thereby cannot be grouped together with decaying interaction strength).
For mean-field models the algorithm may thus not be used (efficiently).

\section{Conclusion \& Outlook}
\label{sec:conclusion}
We have presented a general, hierarchical, and adaptive algorithm for Metropolis Monte Carlo simulations of long-range interacting systems.
The range of possible applications of the algorithm is very broad.
The formulation does not depend on the lattice structure and is thus valid for both general lattice spin models and systems with long-range interacting particles in continuous space as long as the interaction decays sufficiently fast with distance.
In the two applications considered here, viz. the nonconserved long-range Ising model and a Lennard-Jones system in two dimensions we observe runtimes that support an average asymptotic complexity of $O(N\log N)$ (where $N$ is the number of spins or particles) but the existing data for the former may also be described by a power-law with a small exponent.
However, the scenario of a logarithmic scaling seems more likely due to the hierarchical, tree-like nature of the algorithm.
\par
Importantly, our method has small prefactors for the asymptotic scaling of the runtimes, resulting in speed-up factors which exceed $10000$ in relevant physical scenarios.
In a single day, we can perform simulations which before would have taken $\approx 30$ years with any of the established methods, enabling the exploration of parameter ranges that were hitherto not accessible.
\par
Until recently it was only possible to investigate the nonequilibrium properties of the long-range Ising model during quenches to low temperatures where the low acceptance rates allow an efficient simulation via the effective field approach~\cite{christiansen2018}.
The application of the new algorithm is not limited to this scenario, proving very efficient also in case of large acceptance rates, as encountered, e.g., during quenches to the critical temperature.
While here we exemplified our algorithm for the long-range Ising model with nonconserved order parameter and a Lennard-Jones system, our method can easily be applied to other spin and off-lattice systems.
Another promising application is the phase separation in a conserved order parameter simulation of the long-range Ising model where the system evolves at the quench temperature through spin exchanges~\cite{ToBePublished}.
Other nonequilibrium simulation settings where long-range interactions are of interest are field-driven hysteresis~\cite{chakrabarti1999dynamic} and Kibble-Zurek like processes employing a slow quench~\cite{kibble1980some,zurek1985cosmological,zurek1996cosmological}.
Especially interesting is the extension to quantum systems where the Suzuki-Trotter mapping~\cite{Suzuki1976,SUZUKI1993432} of a $d$-dimensional quantum system to the corresponding $(d+1)$-dimensional classical system allows the application of our algorithm, which is designed for general $d$.
Very recently, motivated by D-Wave experiments~\cite{bando2020probing}, Bando and Nishimori~\cite{bando2021simulated} investigated the generalized quantum Kibble-Zurek mechanism in the transverse-field Ising model coupled to an external bath where long-range interactions arise naturally in Trotter direction.
Also of great interest are models where the quantum spins themselves interact via a tunable long-range potential~\cite{jaschke2017critical,puebla2019quantum}, which describe many experimental situations~\cite{kim2010quantum,labuhn2016tunable}.
The algorithm presented here constitutes an important step towards an efficient simulation of the corresponding classical system.

\begin{acknowledgments}
  This project was funded by the Deutsche Forschungsgemeinschaft (DFG, German Research Foundation) under project No.\ 189\,853\,844 -- SFB/TRR 102 (project B04), and the Deutsch-Französische Hochschule (DFH-UFA) through the Doctoral College ``$\mathbb{L}^4$'' under Grant No.\ CDFA-02-07. We further acknowledge support by the Leipzig Graduate School of Natural Sciences ``BuildMoNa''.
  \end{acknowledgments}

\end{document}